\documentclass[aps,prl,twocolumn,superscriptaddress]{revtex4}

\usepackage{mathptmx}
\usepackage{subfigure}
\usepackage{dcolumn}
\usepackage{amsmath,amssymb}
\usepackage{bm}
\usepackage{color}
\usepackage{latexsym}
\usepackage[english]{babel}
\usepackage{times}

\usepackage{psfrag,graphicx} 
\usepackage{epsf}  
\usepackage{amsfonts}
\usepackage{natbib}
\usepackage{epstopdf}
\usepackage{appendix}
\usepackage[colorlinks=true,citecolor=blue,linkcolor=blue]{hyperref}

\begin{document}

\title{A single impurity in an ideal atomic Fermi gas: current understanding and some open problems}

\author{Zhihao Lan}
\email{z.lan@soton.ac.uk}
\affiliation{Mathematical Sciences, University of Southampton, Highfield, Southampton, SO17 1BJ, United Kingdom}

\author{Carlos Lobo}
\affiliation{Mathematical Sciences, University of Southampton, Highfield, Southampton, SO17 1BJ, United Kingdom}

\date{\today}

\renewcommand{\u}{\uparrow}
\renewcommand{\d}{\downarrow}
\newcommand{\fss}{f_0^{\sigma\sigma^\prime}}
\newcommand{\fud}{{f_0^{\u\d}}}
\newcommand{\fuu}{{f_0^{\u\u}}}
\newcommand{\fdd}{{f_0^{\d\d}}}
\newcommand{\bp}{{\bf p}}
\newcommand{\bq}{{\bf q}}
\newcommand{\bP}{{\bf P}}
\newcommand{\bR}{{\bf R}}
\newcommand{\br}{{\bf r}}
\newcommand{\bk}{{\bf k}}
\newcommand{\ms}{{m^*_\sigma}}
\newcommand{\mup}{{m*_\u}}
\newcommand{\mdn}{{m^*_\d}}
\newcommand{\mus}{{\mu_\sigma}}
\newcommand{\muup}{{\mu_\u}}
\newcommand{\mudn}{{\mu_\d}}
\newcommand{\fssp}{{f_{\p \sigma, \p^\prime \sigma^\prime}}}
\newcommand{\ps}{{{\bf p} \sigma}}
\newcommand{\psp}{{{\bf p}^\prime \sigma^\prime}}
\newcommand{\kf}{1/k_Fa}

\begin{abstract}
We briefly review some current theoretical and experimental aspects of the problem of a single spinless impurity in a 3D polarised atomic Fermi gas at zero temperature where the interactions can be tuned using a wide Feshbach resonance. We show that various few-body states in vacuum composed of the impurity and background gas atoms (single impurity, dimer, trimer, tetramer) give rise to corresponding dressed states ({\em polaron}, {\em dimeron},  {\em trimeron}, {\em tetrameron}) in the gas  and inherit many of their characteristics. We study the ground state focussing on the choice of wave function and its properties. We raise a few unsolved problems: whether the polaron and dimeron are really separate branches, what other few-body states might exist, the nature of the groundstate for large numbers of particle-hole pairs and why is the polaron ansatz so good. We then turn to the excited states, and to the calculation of the effective mass. We examine the bounds on the effective mass and raise a conjecture about that of composite quasiparticle states. 
\end{abstract}

\maketitle

\section{Introduction}
What happens when we immerse an impurity in an ideal Fermi gas? This question has a long history, going back at least to the motion of ions in liquid $^3$He and to dilute mixtures of $^3$He in $^4$He \cite{Wolfle}. In cold atomic gases, the impurity as a quasiparticle was first studied in the context of a partially polarised Fermi gas \cite{Lobo2006,ChevyPRA}. The atomic gas case is interesting because of its simplicity and experimental manipulability compared with that of its predecessors: there are only contact interactions between the impurity and the gas atoms which themselves constitute an ideal Fermi gas, as opposed to a strongly interacting system like liquid helium.

Here we will consider the case of an atomic impurity of mass $M$ immersed in a zero temperature gas of another species of mass $m$ which may be the same type of atom in a different spin state or a different type of atom altogether \cite{footnote1} . The impurity interacts with the background gas via a tunable s-wave interaction and there is only one spin state per species. Calling the background atoms ``$\u$" and the impurity ``$\d$", the Hamiltonian is
\begin{equation}
\hat{H}=\sum_\bk \left( \epsilon_\bk \hat{a}^\dagger_{\bk \u} \hat{a}_{\bk \u} + E_\bk \hat{a}^\dagger_{\bk \d} \hat{a}_{\bk \d} \right) + \frac{g_0}{V} \sum_{\bk,\bk^\prime,\bq} \hat{a}^\dagger_{\bk+\bq \u} \hat{a}^\dagger_{\bk^\prime-\bq \d} \hat{a}_{\bk^\prime \d} \hat{a}_{\bk \u} \label{H1}
\end{equation}
with
\begin{equation}
\epsilon_\bk \equiv \frac{\hbar^2 k^2}{2m} \mbox{,    } \, E_\bk \equiv \frac{\hbar^2 k^2}{2M}. \label{H2}
\end{equation}
The gas is in a volume $V$ with periodic boundary conditions; $\hat{a}_{\bk \sigma} \mbox{ and } \hat{a}^\dagger_{\bk \sigma}$ obey the usual anticommutation relations, except if the labels $\u,\d$ refer to different atomic species, in which case they commute.
Also, $g_0$ is the vacuum T-matrix defined in terms of a cutoff $k_c$ (which has to be taken to infinity at the end of the calculation) as
\begin{equation}
\frac{1}{g_0}\equiv \frac{1}{g}-\frac{1}{V} \sum_{k<k_c} \frac{1}{\epsilon_\bk+E_\bk} \mbox{, \, with } g\equiv \frac{2 \pi \hbar^2 a}{m_r} \label{g}
\end{equation}
where $a$ is the s-wave scattering length characterising the interaction between the impurity and the background atoms, and $m_r\equiv mM/(m+M)$ is the reduced mass. This Hamiltonian corresponds to the so-called {\em wide resonance} case where both the scattering length $a$ and interatomic distance between $\u$ atoms are much greater than the effective range characterising the $\u-\d$ interaction. Studies have also been made \cite{YvanNarrow,Massignan,Zhai,Grimm} of the {\em narrow resonance} case where this does not occur and we have to take into account the effective range as a new parameter. There are two dimensionless parameters which control the system: the mass ratio $m/M$ and the ratio of the scattering length to the interatomic distance $k_Fa$ where $k_F$ is the Fermi wave vector of the $\u$ atoms.

This Hamiltonian is not valid for $m/M>13.384$ since above that limit it is known that there are four- \cite{CastinMoraPricoupenko} and three-body Efimov \cite{Efimov} bound states in vacuum (i.e., $\u \u \u \d$ or $\u \u \d$). These require extra parameters beyond $a$ (without them, the energies become cutoff-dependent in the zero-range limit model Eq (\ref{H1})). We assume here that this critical mass ratio continues to play the same role in our case even though it is possible that larger bound states with higher mass ratios exist \cite{footnote2}.

We will confine ourselves to the 3D case and (mostly) the wide resonance, focussing on the nature of the ground and low-lying excited states and raising some unresolved questions about them. Important topics which we will not address for lack of space include: experimental methods, impurities in 1D and 2D, the equation of state of partially polarised Fermi gas and itinerant ferromagnetism. For a good review of these subjects see e.g. \cite{BruunReview}. Perhaps the most important omission is also at the frontier of the field: nonequilibrium quasiparticle properties such as impurity diffusion, mobility and decay.

\section{Groundstate}
What are the few-body low-energy stable states when the density of the background gas $n_\u \equiv k_F^3/6 \pi^2$ is zero? If $a<0$ then the only known state is the single impurity. If $a>0$ then there are at least four known states: i) a single impurity; ii) the $\u \d$ dimer with a wave function of the type $\sum_{\bk} \phi_\bk \hat{a}^\dagger_{\bk \u} \hat{a}^\dagger_{-\bk \d} | vac \rangle$ and an energy $-\hbar^2/2 m_r a^2$; iii) for $8.172 < m/M \lesssim 9.5$ there is a lower energy stable state: an $L=1$ trimer ($\u \u \d$) with a wave function $\sum_{\bp,\bk} \phi_{\bp,\bk} a^\dagger_{\bp+\bk \d} a^\dagger_{-\bp \u} a^\dagger_{-\bk \u} | vac \rangle$ \cite{KartavtsevMalykh}; iv) finally, for $9.5 \lesssim m/M <13.384$, a tetramer ($\u \u \u \d$) has been found to be the groundstate for four atoms \cite{Blume}. Note that, obviously, the dimer cannot decay into a trimer although a trimer could decay into a dimer plus an atom if it were higher in energy (we neglect the deep bound dimer states which are not described by our Hamiltonian \cite{footnote3} ), and the same reasoning applies for the stability of the tetramer.

\begin{figure}[!hbp]
\centering
\includegraphics[width=1.0\columnwidth]{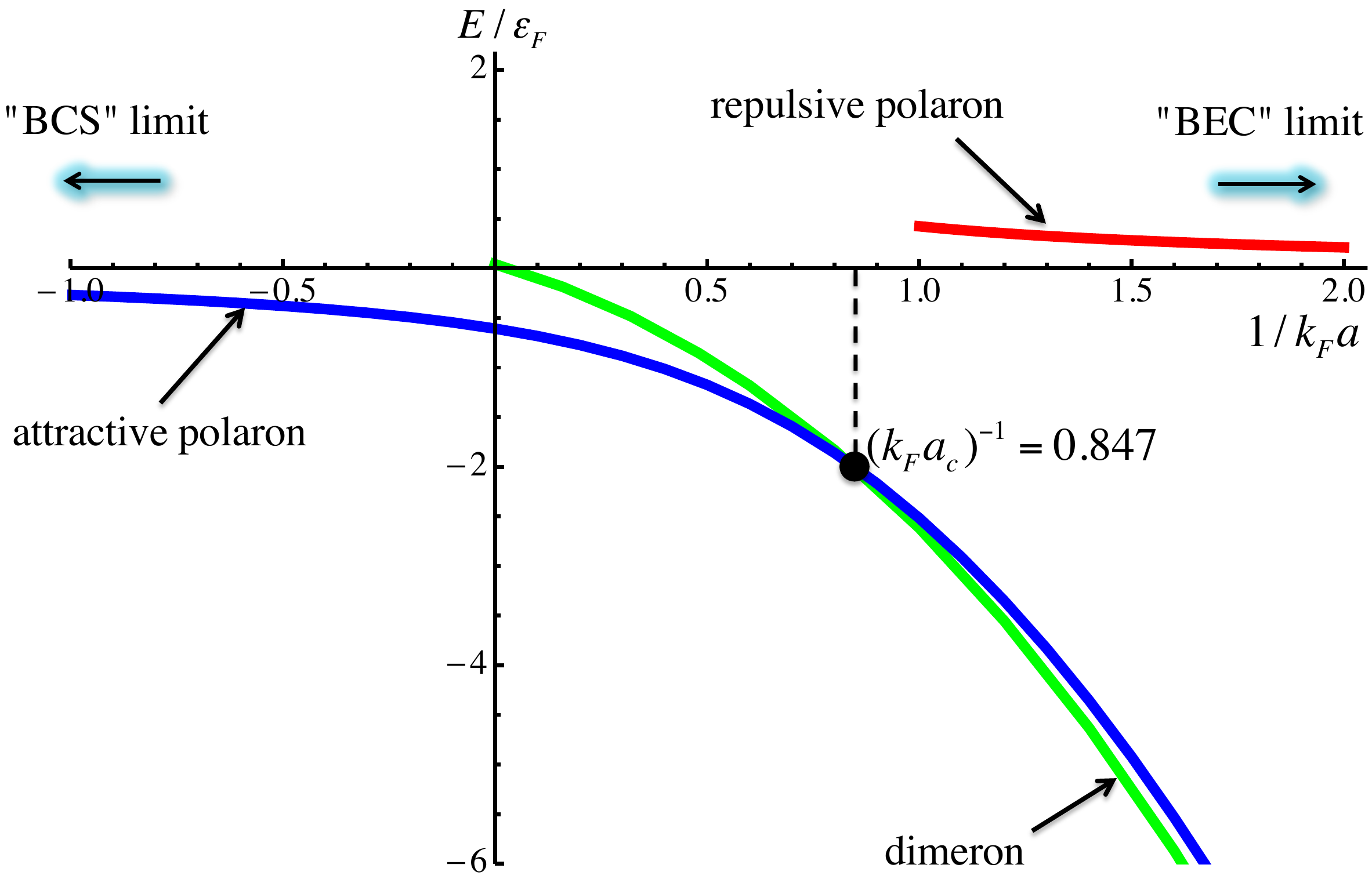} 
\caption{The (grand canonical) energy spectrum of a zero-momentum impurity immersed in a Fermi sea in units of $\epsilon_F$ is a function of $1/k_Fa$. Here we plot the polaron (blue) and dimeron (green) ans\"atze (see Fig. \ref{Fig2}) energies for $m=M$. The repulsive polaron energy (red) is given by $gn_\u$. The branches of the attractive polaron and dimeron cross at $1/k_Fa_c=0.847$ (black dot). The zero of the energy is set at that of the noninteracting system with $N_\u$ atoms (see text).} \label{Fig1}
\end{figure}

Now let us start increasing $n_\u$ (see Fig. \ref{Fig1}). When $a<0$, in the limit $k_F \rightarrow 0$ - the ``BCS" limit - the energy of a single impurity will be reduced from the noninteracting value by the mean field energy $n_\u g(<0)$. Likewise, for $a>0$ - the ``BEC" limit - the single impurity energy is increased by $n_\u g(>0)$; the dimer energy is also increased by the mean field term $n_\u g_{\u {\rm -dimer}}$  where $g_{\u {\rm -dimer}}$ is the same $g$ as in Eq (\ref{g}) but with $a \rightarrow 1.18 a$ (for $M=m$) and $M \rightarrow M+m$ in Eqs (\ref{H2},\ref{g}). The trimer and tetramer will also see their energy shifted although the $\u$-trimer/tetramer scattering lengths are not known at present. All these states are well-defined in the vacuum limit and so we should expect that their adiabatic continuation remains the ground state as we increase $n_\u$ from zero, with the important exception of the single impurity with $a>0$: its energy is positive and so it will be unstable to binding with gas atoms, forming either a dimer or a trimer.

But what happens for stronger interactions beyond mean field? There are some qualitative changes to the vacuum case:
\begin{enumerate} 
\item the few-body system (single impurity, dimer, trimer or tetramer) will interact with the surrounding Fermi gas and give rise to particle-hole fluctuations which reduce the amplitude of the vacuum few-body term in the wave function. E.g. in the single impurity case the quasiparticle residue $Z$ is reduced from unity; nevertheless, over a wide range of parameter space, fluctuations tend to be relatively small (at least for the few-body states explored so far) so that $Z>0$ and the quasiparticle nature inherited from the vacuum few-body system is well-defined \cite{footnote4}. We will use the convention of calling the few-body system dressed by fluctuations a {\em polaron} \cite{Lobo2006,ChevyPRA,CombescotLobo}, {\em dimeron} \cite{CGL,MoraChevy,Zwerger}, {\em trimeron} \cite{Meera3D} or {\em tetrameron}. The dressed state connecting to the single impurity state in vacuum is called the attractive polaron when $a<0$, or repulsive polaron when $a>0$.
\item Other than acting as a source of particle-hole pairs, the most important function of the Fermi sea is to block the momentum of the $\u$ particles from going below $\hbar k_F$.
\item As a consequence of these two effects, the energy in units of $\epsilon_F$ becomes a nonlinear function of $k_Fa$ and the effective mass is shifted from its bare value $M$, $m+M$, $2m+M$ or $3m+M$.
\item Unlike in the vacuum case, when a state is no longer the ground state, it will become a long-lived resonance at best, since it can exchange particles with the gas and decay to the lower energy branch through emission of particles and holes. For example, the repulsive polaron will have a finite lifetime since it can decay into the lower energy branches (see Fig \ref{Fig1}). In 3D the lifetime of the repulsive polaron has only been measured for the narrow resonance, unequal mass case \cite{Grimm} where it was shown that it is surprisingly long-lived even close to resonance: for $1/k_Fa=0.25$, the decay rate $\hbar\Gamma=0.01\epsilon_F$,  which corresponds to a $1/e$ lifetime of about $400\mu s$.  Compared with its energy $E=0.30\epsilon_F$, $\hbar\Gamma/E\sim 0.03 \ll 1$, which shows that the repulsive polaron is a well-defined quasiparticle even deep in the strongly interacting regime..
\end{enumerate}

Thus we see that impurities in a Fermi gas can be best understood as few-body states with particle-hole fluctuations whose vacuum nature is generally preserved  in the gas (at least if they are the groundstate) but whose properties are quantitatively shifted from the vacuum values.

\subsection{Choice of groundstate wave function} 
We shall see that it is convenient not to fix the number of $\u$ atoms but to work with the grand canonical ensemble so that we must minimise $\hat{H}-\mu_\u \hat{N_\u}$.
Before we introduce an impurity into the $T=0$ system we can write down the exact ground state wave function of $N_\u$ atoms which form an ideal Fermi gas:
\begin{equation}
|FS\rangle \equiv \Pi_{|\bk|<k_F} \hat{a}^\dagger_{k \u} | vac \rangle. \label{idealgas}
\end{equation}
The Fermi energy and wave vector are defined in the usual way as $\epsilon_F\equiv \hbar^2 k_F^2/2m(=\mu_\u \mbox{ in our case})$ and $k_F\equiv (6 \pi^2 N_\u/V)^{1/3}$ with $N_\u \equiv \langle \hat{N}_\u \rangle$ so that the density is $n_\u=N_\u/V$.

Considering now the gas plus impurity ground state, the wave function will simply change due to the inclusion of particle-hole fluctuations:
\begin{eqnarray}
\lefteqn{|\Psi \rangle =\left( \phi_0 \hat{a}^\dagger_{\bk=0 \d}  +\sum_{\bk,\bq} \phi_{\bk,\bq} \hat{a}^\dagger_{\bk \u} \hat{a}_{\bq \u} \hat{a}^\dagger_{\bq-\bk \d} +\right.} \label{polaron} \\ &&  \left. \sum_{\bk,\bk^\prime,\bq,\bq^\prime} \phi_{\bk,\bk^\prime,\bq,\bq^\prime} \hat{a}^\dagger_{\bk \u} \hat{a}^\dagger_{\bk^\prime \u} \hat{a}_{\bq \u}\hat{a}_{\bq^\prime \u} \hat{a}^\dagger_{\bq^\prime + \bq-\bk- \bk^\prime \d} + ...\right) |FS \rangle \nonumber
\end{eqnarray}
where the $\phi$s are real coefficients to be determined from the ground state solution and are anti-symmetric with respect to exchanges of particle or hole coordinates. The sums over $\{\bk\},\{\bq\}$ are restricted to be above and below the Fermi surface respectively. Note that the wave function is an eigenstate of total momentum $\bp=0$ \cite{footnote5}.

\begin{figure}[!hbp]
\centering
\includegraphics[width=0.8\columnwidth]{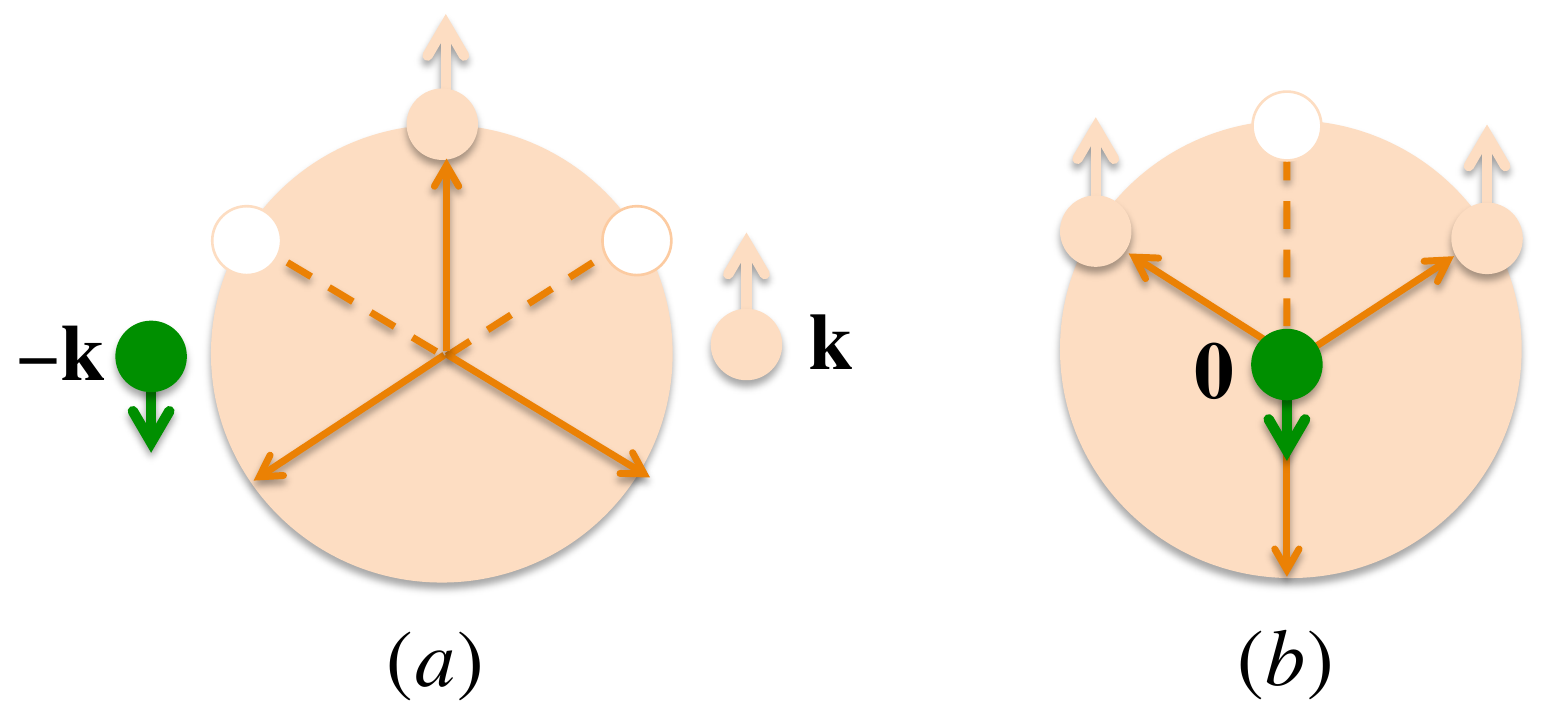}
\caption{Equivalence of Eqs. (\ref{polaron} ) and (\ref{dimeron}) in describing the system. (a) The physics of the first term $\sum_\bk \phi_k \hat{a}^\dagger_{\bk \u} \hat{a}_{-\bk \d}|FS \rangle$ of Eq. (\ref{dimeron}) can be reproduced taking $\phi_0 \sim \phi_{\bk,\bq} \sim 0$, using the third term of Eq. (\ref{polaron}) by fixing two holes and a particle at the Fermi surface so that their total momentum (the sum of the three arrows) is zero, while allowing the remaining atoms at $\bk$ and $-\bk$ to scatter freely as in the dimer state. (b) The first term $\phi_0 \hat{a}^\dagger_{\bk=0 \d}  |FS \rangle$ of Eq. (\ref{polaron}) can be reproduced from the second term of  Eq. (\ref{dimeron}) in the same spirit.} \label{Fig3}
\end{figure}

Although Eq (\ref{polaron}) is in fact an accurate wave function for an impurity in any possible state, it can become a clumsy description of certain ground states which appear as a function of $k_Fa$ and $m/M$. For example, when the dimeron is the groundstate, the $a \rightarrow 0^+$ limiting wave function (the dimer state) $\sim \sum_\bk \phi_k \hat{a}^\dagger_{\bk \u} \hat{a}_{-\bk \d}|FS \rangle$ seems not to be included in Eq (\ref{polaron}). In reality it is there since we can always add zero energy particle-hole pairs which are irrelevant in the thermodynamic limit by taking $\phi_0 \sim \phi_{\bk,\bq} \sim 0$ and, in the third term, fixing two holes and a particle at the Fermi surface so that their total momentum is zero, while allowing the remaining $\u$ and $\d$ atoms to scatter freely as in the dimer state (see Fig. \ref{Fig3}a).

A more convenient representation for a dimeron wave function requires changing the number of $\u$ atoms:
\begin{eqnarray}
\lefteqn{|\Psi \rangle =\left( \sum_{\bk} \phi_\bk \hat{a}^\dagger_{\bk \u} \hat{a}^\dagger_{-\bk \d}   +    \sum_{\bk,\bk^\prime, \bq} \phi_{\bk,\bk^\prime,\bq} \hat{a}^\dagger_{\bk \u} \hat{a}^\dagger_{\bk^\prime \u} \hat{a}_{\bq \u} \hat{a}^\dagger_{\bq-\bk-\bk^\prime \d} +\right.} \label{dimeron} \\ &&  \hspace{-0.5cm} \left.   \sum_{\substack{\bk,\bk^\prime, \bk^{\prime \prime}, \\ \bq,\bq^\prime}}   \phi_{\bk,\bk^\prime,\bk^{\prime \prime},\bq,\bq^\prime} \hat{a}^\dagger_{\bk \u} \hat{a}^\dagger_{\bk^\prime \u} \hat{a}^\dagger_{\bk^{\prime \prime} \u} \hat{a}_{\bq \u}\hat{a}_{\bq^\prime \u} \hat{a}^\dagger_{\bq^\prime + \bq-\bk- \bk^\prime-\bk^{\prime \prime} \d} + ...\right) |FS \rangle \nonumber 
\end{eqnarray}
which can also describe correctly the polaron (see Fig. \ref{Fig3}b).
Likewise, if we were interested in describing a trimeron state, then we would add yet another atom, start with $\sim \hat{a}^\dagger_{\u} \hat{a}^\dagger_{\u} \hat{a}^\dagger_{\d}|vac \rangle$ adding particle-hole fluctuations, and so on for the tetrameron, etc.

\begin{figure}[!hbp]
\centering
\includegraphics[width=0.8\columnwidth]{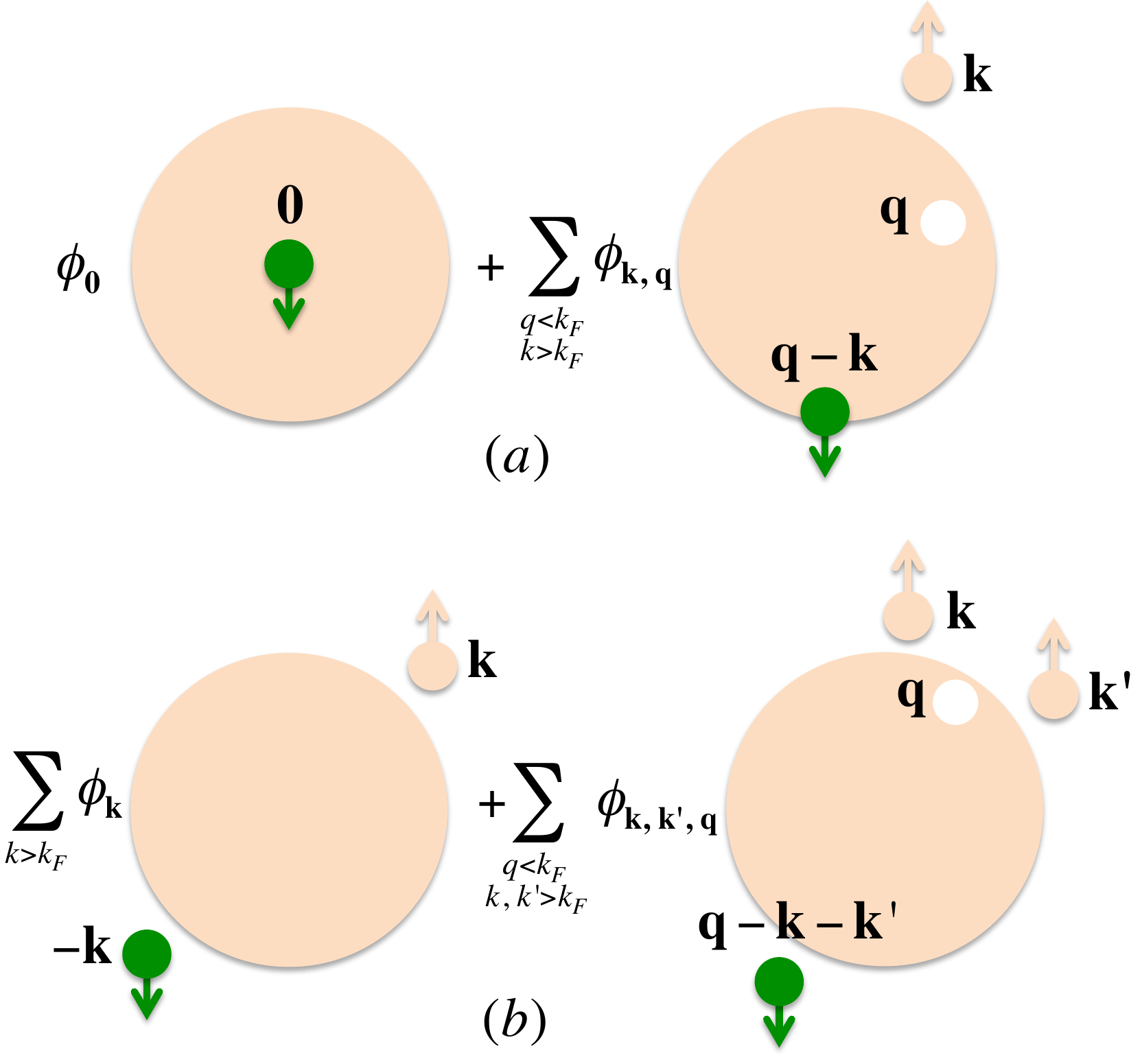}
\caption{Pictorial views of the polaron (a) and dimeron (b) ans\"atze with one particle-hole fluctuation and total momentum $\bp=0$. } \label{Fig2}
\end{figure}

This representation is more convenient in the sense that it reproduces the dimeron physics accurately with the smallest number of particle-hole pairs. It is crucial in practice when we use an approximation scheme and truncate the wave function keeping only a certain number of particle-hole pairs. Then, depending on the order of the truncation, Eqs(\ref{polaron}) and (\ref{dimeron}) can describe very different physics. Truncating Eq (\ref{polaron}) by keeping only one particle-hole pair leaves us with the so-called {\em Chevy} or {\em polaron ansatz}. Doing the same with Eq (\ref{dimeron}) provides us with the {\em dimeron ansatz}  (see Fig. \ref{Fig2}). Surprisingly, keeping only one particle-hole pair is an excellent approximation for the quasiparticle energy even for strong interactions.

\subsection{Ground state properties}
We can now use the two {\em ans\"atze} as variational wave functions, minimising $\langle \hat{H}-\mu \hat{N_\u} - E_0 \rangle$. Setting the zero of the energy at that of the grand canonical energy of the noninteracting ideal Fermi gas with $N_\u$ atoms $-2/5 N_\u \epsilon_F$, we find the two lower curves of Fig. \ref{Fig1} ($m=M$) \cite{footnote6}. We see that, as discussed above, the polaron becomes the groundstate in the BCS limit while the same is true for the dimeron in the BEC limit. Note that Eq (\ref{dimeron}) has $N_\u+1\d$ atoms and so its free energy is $-\hbar^2/2m_ra^2 - \epsilon_F$ when $\kf \rightarrow 0^+$.

Experimentally, various properties of the 3D polaron groundstate have been measured: the energy, the impurity spectral function $A(k,\omega)$ and the quasiparticle residue $Z$ both for wide resonance ($m=M$) \cite{Zwierlein,SalomonScience}, and for narrow resonance (unequal masses) \cite{Grimm}. For the narrow resonance repulsive polaron these quantities have also been measured as well as the lifetime \cite{Grimm}. The techniques used were a combination of rf-spectroscopy and measurement of density profiles. The agreement with the Chevy ansatz for the energy is excellent. For $Z$, there is some disagreement with the experiment in \cite{Zwierlein} but excellent agreement in \cite{Grimm} which uses a different experimental method. Remarkably, there has not been a direct measurement of dimeronic properties so far. In \cite{Grimm}, excited dimeron-hole states were found as predicted \cite{YvanNarrow}, but, for example, there is only indirect evidence of the polaron-to-dimeron transition (see \cite{Zwierlein} where the transition is identified with the vanishing of $Z$). Finally, there has been no measurement at all of trimeron or tetrameron states.
Overall, the main experimental conclusion is that the polaron ansatz is in excellent agreement with experiments and it reproduces with substantial accuracy the results of more sophisticated theoretical calculations \cite{CombescotGiraud,ProkofevSvistunovPolaron,Houcke3D}.

\subsection{Some open problems}

\subsubsection{Are the dimeron and polaron really separate energy branches?}
Edwards \cite{Edwards} has raised the question of whether the polaron and dimeron energy branches are really separate or one and the same (i.e. is the groundstate wave function continuous across the polaron-to-dimeron transition at $k_Fa\sim 1$?). All calculations indicate so far that the wave functions are different when their energies cross. They show a slope discontinuity of the energy as a function of $1/k_Fa$ of $0.75 \epsilon_F$ and a discontinuous change of $Z$ from $\sim 0.3$ (on the polaron side) to exactly zero on the dimeron side \cite{Zwerger, footnote7}. Experimentally however, $Z\rightarrow 0$ around $k_Fa \sim1$ \cite{Zwierlein} but it seems to do it continuously. Likewise, there is no evidence for a discontinuity in the slope of the energy in any experiment so far.

\begin{figure}[!hbp]
\centering
\includegraphics[width=1.0\columnwidth]{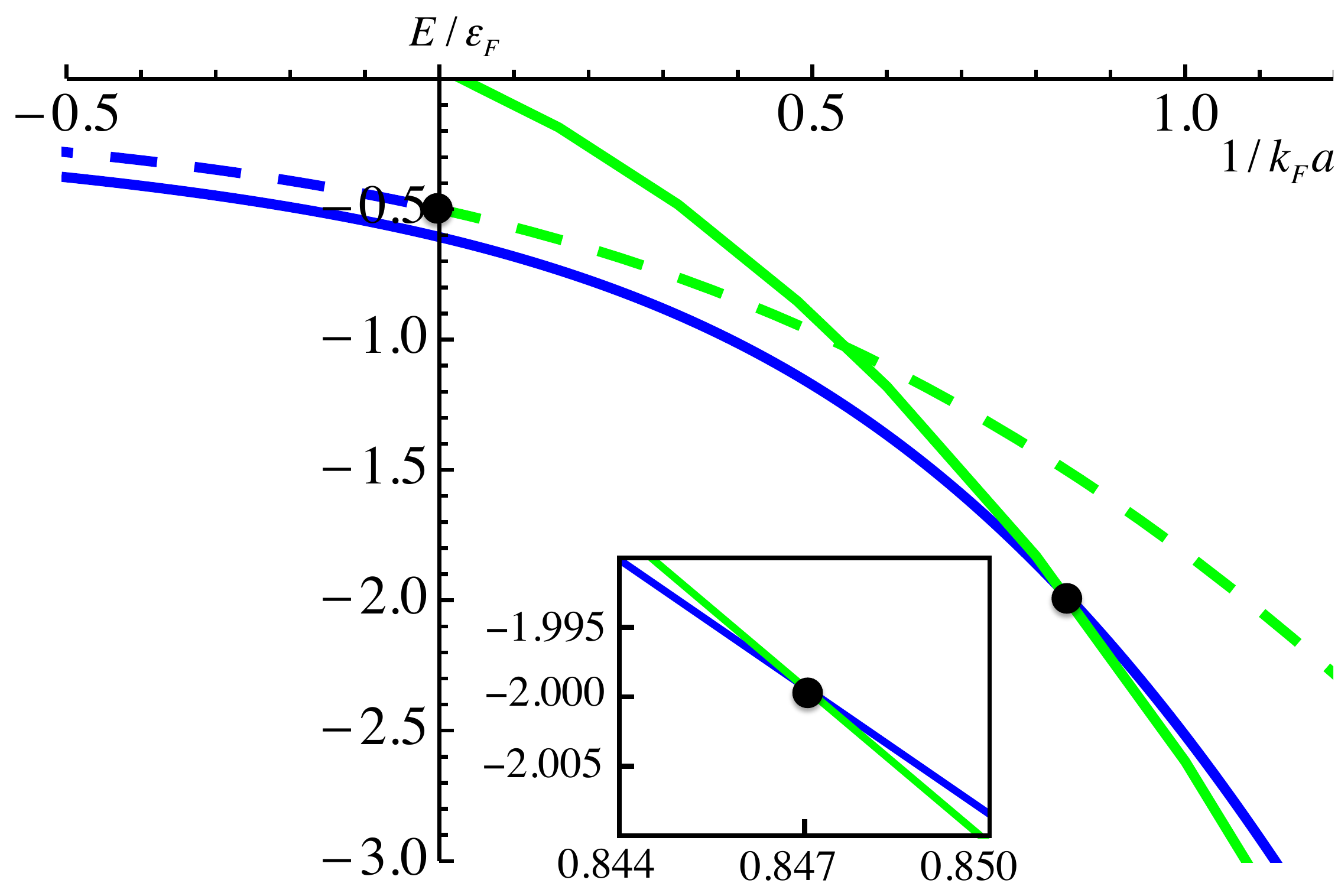}
\caption{Shift of the polaron-to-dimeron transition (black dots) from $\kf_c=0.847$ for $m=M$ to $\kf_c=0$ in the limit $m/M\rightarrow 0$ where the  transition becomes continuous. The solid curves are for $M=m$ while the dashed is for $m/M\rightarrow 0$ (blue for polaron and green for dimeron). } \label{Fig4}
\end{figure}

But, playing devil's advocate, we could argue that calculations are always approximate, whether variational or Monte Carlo, and the experiments are not precise enough to tell either way. Is there a more definitive theoretical argument? Barring that, can we at least make a plausibility argument? A simple idea is that the two branches have different statistics i.e., if we were to transport adiabatically two localised polarons so that they exchanged positions, the many-body wave function would change sign reflecting the fermionic statistics of the polaron, whereas exchanging two localised dimerons would return it to the same value. This is exactly true in the deep BCS (for the polaron) and BEC (dimeron) limits. The difficulty is that adiabatic transport might not be possible in the presence of a free (i.e. ungapped) Fermi surface since there is a finite density of states of zero energy particle-hole excitations so the best we could hope for would be an approximate statement. Nevertheless, if we accept some version of this argument, we see that the wave function cannot continuously change its discrete permutation symmetry upon an infinitesimal change in $1/k_Fa$ and so the two branches must be separate.

However, even if we accept that there is a discontinuity for $m=M$, the problem becomes more acute in the $m/M\rightarrow 0$ limit. Then, if we treat the impurity as fixed, which is often done in the literature to find the energy of the infinitely heavy polaron \cite{CombescotLobo}, we find, surprisingly, that the polaron-to-dimeron transition is continuous in the thermodynamic limit and occurs at unitarity, corresponding simply to the appearance of a bound state in the potential of the impurity \cite{KohnMajumdar}. Is this a smooth transition at the critical $1/k_Fa_c$ as a function of $m/M$:
\begin{eqnarray}
&&\left(\frac{dE_{\rm polaron}}{d (\kf)}-\frac{dE_{\rm dimeron}}{d (\kf)} \right)_{1/k_Fa_c} \propto \left( \frac{m}{M} \right)^\alpha \\
&& \left(Z_{\rm polaron} \right)_{1/k_Fa_c} \propto \left( \frac{m}{M} \right)^\beta
\end{eqnarray}
with $\alpha, \beta >0$? Or does it happen discontinuously?

And what happened to the arguments given above? For large mass it is known that $Z \rightarrow 0$ as $m/M\rightarrow 0$ \cite{Rosch,PerturbativeCastin} for the polaron so that both branches would have $Z=0$. Also, for small $m/M$, given that the Fermi sea is ungapped, Anderson's orthogonality catastrophe \cite{Anderson} leads to loss of coherence on a scale $1/k_F$, so that it is likely that a statistical sign cannot be defined since the wave function would not return to itself due to the creation of a large number of particle-hole pairs.

\subsubsection{What few-body bound states with a single impurity and any number of identical fermions are stable in vacuum for a given $m/M$ and $a>0$?}
In a vacuum, a larger bound state can decay into a smaller one if its energy is higher. For example, a trimer could decay into a dimer plus an atom. The opposite process cannot obviously occur in the vacuum but can certainly happen in a degenerate gas: the dimer could form a trimer by combining with a gas atom. This means that the groundstates in the gas are the lowest energy ones regardless of their size. The question then is: what are the true lowest energy states in vacuum? These will give rise to groundstate branches at finite gas density.
As we saw, for $8.172< m/M \lesssim 9.5$ a $\d \u \u$ trimer is stable in vacuum leading to a trimeron phase in the gas. For $9.5 \lesssim m/M < 13.384$ a $\d \u \u \u$ tetramer is thought to be the groundstate in vacuum leading to a corresponding tetrameron phase (as before we will not discuss recombination to deep bound states, assuming that it is small for non-Efimovian states in the sense of \cite{footnote3}). For $m/M>13.384$, Efimov states appear which lead to rapid loss.

Assuming that $m/M<13.384$ is the lower limit for the appearance of Efimov states, can there be other larger, stable, non-Efimovian few-body states (pentamers, hexamers and so on) below that limit? These have not been investigated to our knowledge since they are computationally unwieldy. They would still be described by the wide resonance Hamiltonian and could lead to corresponding phases in the gas. Perhaps, from the theoretical point of view, it is more interesting to ask the question: could we prove the existence of an upper limit (with Fermi sea present or in vacuum) to the number of fermions in a non-Efimovian bound state with a single impurity for given $m/M$, $a$? Intuitively the fermions would like to get closer to the impurity to lower their interaction energy. However, Pauli blocking would then increase their kinetic energy and the problem would get worse for larger number of fermions.

\subsubsection{What is the asymptotic wave function for large particle-hole numbers?}
We have discussed mainly the first few terms of the wave function of the impurity because they carry most of the probability. So far there has been no study of the ``tail" of the wave function - the asymptotic form for large number $s$ of particle-hole pairs. The theoretical interest here is the connection with the problem of the coherence of the impurity in a Fermi gas. There is some hope of being able to do this analytically since, for very large $s$, we are dealing with a gas of noninteracting particles which only scatter with the single impurity so that it is likely that correlations between different particles and holes are entirely lost. In this case we might be able to invoke a single mean-field $n_{p \u}$, the atomic density, to characterise the gas, for example, by solving self-consistently the scattering of a single particle with the impurity using $n_{p \u}$ as a Fermi factor as in BCS theory, which would then be itself a function of the scattering solution summed over all particles. As mentioned above, the exact solution for the infinitely heavy impurity (fixed scatterer) is straightforward and provides a limiting case.

\subsubsection{Why is the Chevy ansatz so good?}
The Chevy ansatz seems to be extremely good even for large scattering length where we would expect strong particle-hole correlations, particularly for the energy and $Z$ \cite{Houcke3D}. Why? Does the particle-hole pair expansion of the wave function have a hidden small parameter? And when does it fail (if ever)? We would like to have a more quantitative control of the error when truncating the expansion at a certain order. There are two important clues: first, the contact wide-resonance interaction has no momentum dependence so that, near $k_Fa \gg 1$, the only momentum scale is $\hbar k_F$. Since hole momenta $\{q\}<k_F$ and the particle momenta $\{k\}>k_F$, we can expand denominators occurring in the expression for the energy of the polaron in powers of $q/k$ and the $q=0$ term dominates \cite{Recati,CombescotGiraud}. Second: contributions to the polaron energy which differ by exchange of hole momenta cancel out due to hole antisymmetry (to lowest order in $q/k$) \cite{CombescotGiraud}. This makes the corrections to the Chevy ansatz energy when adding another particle-hole pair very small. Are these two effects enough to show that terms with larger number of particle-hole pairs have smaller contributions and can we estimate the error in various quantities when truncating to a certain order in particle-hole pair number? Also, how would this error vary with $m/M$ and $\kf$?

\section{Excited states}

The exact excited states of the Hamiltonian Eq (1) are in general quite complex. But it is physically more relevant to characterise the long-lived low-energy resonances. We can see them as either resonances due to the coupling of a bare state to a continuum \cite{YvanNarrow} or as excitations of the dressed impurity (polaron, dimeron, etc) which decay due to scattering of particle-hole pairs so that their energy is complex. Following the Fermi liquid approach, we assume that there is an adiabatic process connecting the excited states of the uncoupled system with those of the coupled system so that they can be classified using the same quantum numbers.

In the BCS limit there are no interactions and the bare states are uncoupled. These are simply the momentum states of the impurity with energy $p^2/2M$ and the particle and hole excitations of the ideal Fermi gas. As we turn on the interactions, at low momenta (to order $p^2$), the real part $E_{\bp}$ of the excited state energy is shifted with respect to the groundstate energy $E_0$ and the shift is characterised by a single parameter, the effective mass $M^*$, becoming $E_\bp=E_0+p^2/2M^*$. The imaginary part is $\propto p^4$ and $\ll E_\bp-E_0$ \cite{LanBruunLobo} for small $p$. I.e. the excitation - or {\em quasiparticle} - is long-lived and has a renormalised mass. In contrast, the particle-hole excitation energy remains the same to $O(1/N_\u)$.

In the BEC limit the repulsive polaron is always an excited state since it can decay as discussed above. We have to be more careful with the dimeron, trimeron and tetrameron since they are adiabatically connected to {\em bound} states, i.e. they cannot be obtained by assuming $a=0$ but rather by taking a limiting process $a \rightarrow 0^+$. The dressed excited states will inherit the quantum numbers of the ideal gas, plus the internal (e.g. higher angular momentum \cite{YvanNarrow}) and external ones of the bound state. The energy of the bound state is again of the form $E_\bp=E_0+p^2/2M_{\rm b}^*$ where $M_{\rm b}^*$ is the effective mass of the dimeron and so on.

Experiments measuring $M^*$ at $\kf=0$, use collective modes \cite{SalomonPRL} and density profile analysis fitted to equations of state \cite{SalomonScience,Shin}. They are in good agreement with the polaron ansatz, Monte Carlo methods\cite{CombescotLobo,ProkofevSvistunovPolaron,Houcke3D} and with an ansatz which involves {\em two} particle-hole pairs \cite{CGL}. Interestingly, the polaron ansatz starts to differ significantly from the other theoretical methods as we increase $\kf$ towards the BEC side.

The calculation of $M^*$ using a variational method raises some delicate points. The trial wave function used is an eigenstate of total momentum $\bp \ll \hbar k_F$ and its minimised energy $E_\bp$ is fitted to the formula above to extract $M^*$. However, if we were to use Eq (\ref{polaron}) generalised to finite momentum for example we would find that $M^*=\infty$! This is because the groundstate at finite $\bp$ corresponds to a polaron at zero momentum plus zero energy particle-hole pairs which carry the momentum (or an infinitesimal shift of the Fermi sea as in \cite{footnote5}) so that $E_\bp=E_0$. In practice, the trial wave function is truncated and this leads to a finite $M^*$ since at low energies the polaron can no longer decay to zero momentum in this truncated Hilbert space. This raises the question of whether the variational method is a well defined procedure for finding the effective mass. As far as we know it has not been shown that this corresponds to calculating the real part of the complex excitation energy discussed above. Presumably this method works when the imaginary part is small. Even if we accept this, some care must be taken in the minimisation procedure since, in the thermodynamic limit, it is always possible to find $\bp_c$ such that $E_{\bp_c}^s=E_0^{s-1}$ where $s$ is the number of particle-hole pairs in the wave function (we assume that $E_0^s<E_0^{s-1}$ for all $s$). Then we see that a wave function with $s$ particle-hole pairs and total momentum $\bp_c$ could describe either a polaron with momentum $\bp_c$ or one at rest (with $s-1$ pairs) plus a single particle-hole pair carrying $\bp_c$ but with zero energy (which is always possible in the presence of a Fermi sea). So a correct minimisation of this wave function yields an energy $E_0^s+\bp^2/2M^*$ ($p<p_c$) or $E_0^{s-1}$ ($p>p_c$). This also explains why the use of the $s=1$ (Chevy) ansatz gives us the correct effective mass over such a wide range of momentum: for this case, $p_c/p_F \sim 0.84$ (at unitarity and $m=M$). For $s=2$ however, $p_c/p_F \sim 0.1$ and we must restrict ourselves to lower momenta. At any rate, a necessary condition for the variational method to yield $M^*$ is that $E_0^s<E_0^{s-1}$ for all $s$ so that there is some open set where $d^2E_\bp/dp^2>0$ \cite{CastinMovingPolaron}.

\subsubsection{What are the constraints on the effective mass?}
For the polaron, we must have $M^*>0$ since otherwise the groundstate would be unstable. But are there any other constraints? We can place a simple bound on $E_\bp-E_0$ using a variational argument (keeping in mind the caveats above) \cite{Tony}: consider a trial wave function for the polaron with momentum $\bp$: $\Psi_\bp=\exp(i \bp \cdot \br/\hbar) \Psi_0$ where $\br$ is the coordinate of the impurity and $\Psi_0$ is the zero momentum many-body groundstate with energy $E_0$ which depends on the coordinates of all the atoms including the impurity. $\Psi_\bp$ attempts to give the impurity a finite momentum $\bp$ and is an exact eigenstate of the noninteracting system. We write the full Hamiltonian separating explicitly the kinetic energy operator of the impurity as $\hat{H}=\hat{\bp}_{\rm imp}^2/2M + \hat{H}^\prime$. We now evaluate the energy of the trial wave function as:
\begin{eqnarray}
E_\bp &\leq& \langle \Psi_\bp | \hat{\bp}_{\rm imp}^2/2M + \hat{H}^\prime | \Psi_\bp \rangle \nonumber \\ &=&\frac{p^2}{2M} + 2 \frac{\bp}{2M} \cdot \langle \Psi_0 | \hat{\bp}_{\rm imp} | \Psi_0 \rangle +\langle \Psi_0 | \hat{\bp}^2_{\rm imp}/2M + \hat{H}^\prime| \Psi_0 \rangle \nonumber \\ &=&\frac{p^2}{2M} +E_0
\end{eqnarray}
since the $\langle \hat{\bp}_{\rm imp} \rangle=0$ in the groundstate \cite{footnote9} . By assumption, $E_\bp-E_0=p^2/2M^*$ and that it follows immediately that
\begin{equation}
\frac{M^*}{M}\geq 1. \label{bound}
\end{equation}
To date, all measurements and calculations of $M^*$ obey this inequality (in the groundstate). 

Note that this argument could also be used for the dimeron and other states and we would obtain the equivalent bound $M_b^*/M \geq 1$ where $M_{\rm b}^*$ is the  effective mass of the bound state. However this is not a very tight bound. E.g. for the dimeron, $M_{\rm b}^* \rightarrow M_{\rm b}=M+m$ in the BEC limit ($M_{\rm b}$ being the bare mass) so that the bound trivially states that $m$ is positive. Is there a better bound for dimerons, trimerons and tetramerons? We conjecture that the dressing always increases the mass, even for composite particles:
\begin{equation}
\frac{M^*_{\rm b}}{M_{\rm b}} \geq 1 ?
\end{equation}
All calculations of the effective mass of the dimeron so far are consistent with this simple generalisation \cite{YvanNarrow,CGL}.

Acknowledgments: we thank Y. Castin for helpful discussions and EPSRC for support through grant EP/I018514/1.


\begin{thebibliography}{99}
\bibitem{footnote1} In the case of different spin states of the same atom, the main difference would be that the wave function would not be necessarily an eigenstate of the total spin $S_z$ (we could have a population of impurities in a superposition of $\u$ and $\d$). Impurity states of this type would be important when studying e.g. transverse spin transport.
\bibitem{footnote2} In \cite{Meera3D} it was seen numerically that the 13.6 mass ratio still led to a cutoff dependence of the energy for the $\u \u \d$ case even when the heavy atoms were blocked due to the presence of a Fermi sea.
\bibitem{footnote4} It is well-known that, in 2D and 3D, $Z>0$ for single impurities in Fermi liquids \cite{Rosch}.
\bibitem{footnote5}This is in principle necessary since we would always be able to lower the energy by going to a frame of reference where $\bp=0$ by Galilean invariance. In our case this is unnecessary in the thermodynamic limit: it would correspond to going to a frame with velocity $\propto 1/N_\u$ or, interpreted in the original frame, to a creation of $O(1)$ numbers of particle-hole pairs of zero energy at the Fermi surface which would cancel the momentum of the state. However, sometimes it can be convenient to exploit this freedom and minimise the energy as a function of $\bp$ with the small shift of the Fermi sea being implied rather than explicitly accounted for in the wave function. This is especially important in cases where the effective mass becomes negative, so that increasing the total momentum tends to reduce the energy as in \cite{Meera3D}. However we shall not consider these states here.
\bibitem{footnote6} The repulsive polaron energy should not be obtained within a variational ansatz but rather with a perturbative calculation.
\bibitem{footnote7}This raises the question of whether it is possible to define dimeron, trimeron or tetrameron residues and if these could be measured experimentally.
\bibitem{footnote9} This argument explicitly excludes cases where the impurity has finite momentum in the groundstate. See \cite{Meera3D} for such a case. It also excludes the repulsive polaron since it is not the groundstate.
\end{thebibliography}
\end{document}